\documentstyle[12pt]{article}

\def\ben{\begin{enumerate}}
\def\een{\end{enumerate}}
\def\beq{\begin{equation}}
\def\eeq{\end{equation}}
\def\bea{\begin{eqnarray}}
\def\eea{\end{eqnarray}}
\def\beann{\begin{eqnarray*}}
\def\eeann{\end{eqnarray*}}
\def\beasn{\begin{sneqnarray}}
\def\eeasn{\end{sneqnarray}}
\def\Pf{\mbox{Pf\,\,}}
\def\nobiblabels{\def\@lbibitem[##1]##2{\@bibitem{##2}}}
\makeatother
\def\tabaddress#1{{\small\it\begin{tabular}[t]{c}#1\\[1.2ex]\end{tabular}}}

\begin{document}
\thispagestyle{empty}
\title{Dirac bracket and Nambu structures}
\author{\sc
J. Antonio Garc\'ia\thanks{ E-mail: garcia@nucleares.unam.mx}\, and
\sc
Rafael Cruz-Alvarez\thanks{ E-mail: rafael.cruz@correo.nucleares.unam.mx}\,
\\
\def\baselinestretch{1.2}
\tabaddress{ Instituto de Ciencias Nucleares,\\
Universidad Nacional Aut\'onoma de M\'exico,\\
Apartado Postal 70-543,\\
Ciudad de M\'exico, 04510 M\'exico.} 
}
\date{}

\def\baselinestretch{1.2}
\nobiblabels
\def\theequation{\thesection.\arabic{equation}}

\maketitle

\begin{abstract}
A relation between the Dirac bracket (DB) and Nambu bracket (NB) is presented. 
The  Nambu bracket can be related with Dirac bracket if we can write the DB as a generalized
Poisson structure.  The NB associated with DB have
all the standard properties of the original DB. When the 
dimension of the phase space is $s+2$ where $s$ is the number of second class
constraints, the associated Nambu structure has $s+2$ entries and reduces
to the Dirac bracket when $s$ of its entries are fixed to be the second class
constraints. In general, when
the dimension of phase space is $d=r+s$ a new Nambu structure that describes
correctly the constrained dynamics can also be constructed but in thsi case addicional conditidionts are requiered.  
In that case the associated NB corresponds to a ``Dirac-Nambu'' 
bracket with $r$ entries.
\end{abstract}
\vfill
\clearpage

\section{Introduction}

The construction of Hamiltonian structures from a given set of  first order differential equations and its associated constants of motion in Hamiltonian mechanics has been analyzed \cite{Hojman1}. 
We will take this reference as an interesting starting point to construct a new form of the DB. 
In
this reference, it is also suggested that the Dirac bracket can be viewed as of generalized Poisson structure. In the following we will show that in fact it is possible to construct a new form of Dirac bracket that can be compared and related with analogous Nambu structures. 


The purpose of this note is
to show that these DB can be formulated as a singular Poisson
structure with remarkable similarities to the analysis presented also in
\cite{HDR}. 

To construct Poisson
structures it is necessary to solve a set of conditions where perhaps the most difficult one is the Jacobi Identity. A solution to these conditions  can be constructed in terms of 
the antisymmetric
Levi-Civita tensor and some Casimir functions of the corresponding
bracket. The solution also involves an overall arbitrary factor. 
We will show that the DB can be constructed in an analogous way but the role of the Casimir functions are now played by the set second class constraints obtained in  Dirac analysis of Hamiltonian
constrained theories  \cite{Dirac},\cite{HT}. 
 If the dimension of the phase space is $n$ and the number of
second class constraints is $s$ 
the Dirac symplectic structure can be constructed from an antisymmetric tensor that can be recursively constructed from products of the symplectic form of the ambient phase space. The overall arbitrary factor can then be fixed to be
proportional to the inverse of the Pfaffian of the antisymmetric matrix
constructed from the Poisson brackets of the second class constraints.
Finally, we analyze some of the consequences that can be deduced
from this new formulation of the Dirac symplectic structure. 
In particular, it is now possible to compare this relevant structure
with an alternative formulation of Hamiltonian dynamics based on
the Nambu brackets \cite{Nambu}, \cite{FP}, \cite{Takhtajan}.  A previous approach to relate the NB and DB was presented in \cite{CZ}. We expect that our work can be  complementary and can be used to elucidate many points of previous construction.

In the present work we will just use some concepts that we develop in an informal way with the aim to relate the corresponding brackets in a useful form that can be  applied to other instances.

\section{Poisson structures}

A  Poisson 
manifold is a smooth manifold equipped with a Poisson bracket 
satisfying the skew-symmetry condition, Leibniz rule and Jacobi
identity (see for instance \cite{HT})
\beq
\{A,B\}=-\{B,A\}, 
\label{antisym1}
\eeq
\beq
\{A,BC\}=\{A,B\}C+B\{A,C\},
\eeq
\beq
\{A,\{B,C\}\}+\{C,\{A,B\}\}+\{B,\{C,A\}\}=0.
\label{jacobi1}
\eeq
In coordinates this operation can be realized as
\beq
\{A(z),B(z)\}=\frac{\partial A}{\partial z^i}\sigma^{ij}
\frac{\partial A}{\partial z^j}.
\label{def-bra}
\eeq
where the phase space coordinates are $z^i, i=1...n$ and
$\sigma^{ij}(z)$ is an antisymmetric matrix that fulfills the
conditions
\beq
\sigma^{ij}=-\sigma^{ji}, 
\label{antisym}
\eeq
\beq
\sigma^{ij}\partial_{k}\sigma^{kl}+
\sigma^{li}\partial_{k}\sigma^{kj}+
\sigma^{jl}\partial_{k}\sigma^{ki}=0, \quad \mbox{Jacobi identity}
\label{jacobi}
\eeq
where $\partial_{k}$ denotes the partial derivative with respect to
$z^k$. These conditions are a consequence of the properties 
(\ref{antisym1}) and (\ref{jacobi1}). 
Of course,
the standard formulation of Hamiltonian mechanics is a particular case
of this general construction when $z^i=(q^a,p_b)$ and
\beq
\{z^i,z_j\}=\sigma^{ij}, \quad \sigma^{ij}=\pmatrix{0 & 1\cr -1 & 0}.
\label{standard}
\eeq
This standard structure satisfies in addition to (\ref{antisym},
\ref{jacobi}) the regularity condition
\beq
\det\sigma\not=0.
\label{regular}
\eeq
These conditions are in some sense ``kinematical'' because they do not
depend on the particular dynamics described by some Hamiltonian
function $H(z)$. 

The question about if  a set of first order differential equations
\beq
{\dot z}^i=f^i(z^j), \quad i,j=1...N
\label{system}
\eeq
can be written as Hamiltonian equations for some Hamiltonian function $H$ and
some Poisson bracket structure $\sigma^{ij}(z)$ 
can be resolved in two steps. 
First we need a symplectic two form
that satisfies the properties (\ref{antisym},\ref{jacobi}) and second
we must find a
Hamiltonian function that reproduces the dynamics of the system 
(\ref{system}) through Hamiltonian
equations of motion.  
Let us recall some of the basic
ideas of this construction that are  relevant in what follows.   

To find a Hamiltonian formulation of (\ref{system}) for a
Hamiltonian function $H(z)$ and a two-form $\sigma^{ij}$ that satisfies
the conditions (\ref{antisym},\ref{jacobi}), in addition we impose
\beq
\sigma^{ij}\frac{\partial H}{\partial z^j}=f^i.
\label{dyn-cond}
\eeq
In the particular case when the  condition (\ref{regular}) is
fulfilled, the symplectic 
structure associated with the Poisson
bracket (\ref{def-bra}) is regular. 
This last condition cannot always be met. In particular the
Dirac bracket does not fulfill this condition
whatsoever the dimension of the phase space is. 

Given the set of
equations (\ref{system}) a solution to the
conditions (\ref{antisym},\ref{jacobi}) can be found in the form \cite{HDR}
\beq
\sigma^{ij}=\mu(z)\epsilon^{iji_1....i_{n-2}}\frac{\partial C_1}
{\partial z^{i_1}} 
\frac{\partial C_2}{\partial z^{i_2}}....\frac{\partial C_{n-2}}
{\partial z^{i_{n-2}}}
\label{bra-unconstrained}
\eeq
where $\mu(z)$ is an arbitrary function,  $\epsilon^{iji_1....i_{n-2}}$
is the Levi-Civita antisymmetric tensor and $C_1....C_{n-2}$ is a set of
independent Casimir functions for this bracket structure. 
Notice that this construction is independent of
the dynamics described by (\ref{system}) and depends only on the antisymmetry
property (\ref{antisym}) and Jacobi identity (\ref{jacobi}). 

Now to construct a Hamiltonian
formulation of (\ref{system}) we must use the condition
(\ref{dyn-cond}) 
\beq
\mu(z)\epsilon^{iji_1....i_{n-2}}\frac{\partial C_1}
{\partial z^{i_1}} 
\frac{\partial C_2}{\partial z^{i_2}}....\frac{\partial C_{n-2}}
{\partial z^{i_{n-2}}}\frac{\partial H}{\partial z^j}=f^i
\eeq
From here we solve for some Hamiltonian function $H(z)$ and some
function $\mu(z)$. The evolution equations are then
\beq
\frac{dF}{dt}=\{F,H\}=\partial_i F\sigma^{ij}\partial_j H.
\label{evol}
\eeq
Notice that $H(z)$ is an integral of motion associated to the 
system {\it integrable} system (\ref{system}).
The equations (\ref{evol}) implies that the Casimir functions
$C_\alpha$ 
are in addition to $H$, $n-2$ integrals of motion associated to the 
original system (\ref{system}). Of course it is a very difficult task
to construct Hamiltonian structures in that way because we need a set of
$N-1$ integrals of motion for the given system of $n$
differential equations. A beautiful ansatz to construct this type of structures
from the symmetries of the original system is presented in \cite{Hojman1}. We
refer the reader to this reference for details.

\section{The Dirac bracket}

We will present here a new form of the DB that can be used to relate it with a NB.
For simplicity let us start with a constrained theory, where only second 
class constraints are present, in a phase space defined by the variables 
$z^i=(q^a,p_a)$ with the standard Poisson structure (\ref{standard}).
The Extended Hamiltonian is \cite{HT}
\beq
H_E=H_c(z)+\lambda^\alpha \chi_\alpha(z), \quad i=1...n, \quad \alpha=1...s
\eeq
where $H_c$ is the canonical Hamiltonian, $\lambda^\alpha$ are
Lagrange multipliers and $\chi_\alpha(z)$ is the set of all the 
second class constraints of the theory and $\alpha$ runs over the total set of second class constraints $s$. The Dirac
consistence conditions are satisfied by the canonical Hamiltonian
$H_c$ and the set of constraints $\chi_\alpha$. As all the constraints
are second class the matrix defined by
\beq
\{\chi_\alpha,\chi_\beta\}=C_{\alpha\beta}(z^a)
\label{C}
\eeq
is invertible. We denote the inverse by $C^{\alpha\beta}$. As a
consequence, all the Lagrange multipliers $\lambda_\alpha$ can be obtained as functions
of the phase space. The standard Dirac analysis of this
type of systems conduces to the Dirac bracket \cite{HT}
\beq
\{F(z),G(z)\}^*=\frac{\partial F}{\partial z^i}{\sigma^*}^{ij}
\frac{\partial F}{\partial z^j}, 
\label{Dirac-bra}
\eeq
where the symplectic structure is
\beq
{\sigma^*}^{ij}=\sigma^{ij}-
\sigma^{ik}\frac{\partial \chi_\alpha}{\partial z^k}C^{\alpha\beta}
\frac{\partial \chi_\beta}{\partial z^l}{\sigma}^{lj}.
\label{dirac-sym}
\eeq
This Dirac symplectic structure ${\sigma^*}^{ij}$ satisfies the
fundamental relations (\ref{antisym},\ref{jacobi}) and is singular 
\beq
\det {\sigma^*}^{ij}=0
\eeq
In addition,  the DB satisfy the properties
\beq
\{\chi_\alpha,F\}^*=0
\label{prop1}
\eeq
for $F(z)$ arbitrary. 

If we redefine the constraint surface by
\beq
{\bar\chi}_\alpha=M^\beta_\alpha (z)\chi_\beta
\label{changeM}
\eeq
the Dirac structure is invariant on the constraint surface
\beq
{\bar{\sigma^*}}^{ij}\approx {\sigma^*}^{ij}
\label{prop2}
\eeq
where $\approx$ denotes Dirac weak equality.

Sometimes it is easy to construct the Dirac bracket in the reduced
phase space by enforcing the second class constraints into the
action \cite{FJ}. Then we can read the bracket structure from the kinetic 
term in the deformed reduced action.

\section{The Dirac bracket as a Generalized Hamiltonian structure}

In this section we will show an interesting new form of the Dirac bracket
(\ref{dirac-sym}) can be written as
\beq
{\sigma^*}^{ij}_p=\mu(z)\eta^{iji_1...i_{s-1}i_s}\partial_{i_1}\chi_1
\partial_{i_2}\chi_2...\partial_{i_{s-1}}\chi_{s-1}\partial_{i_s}\chi_s
\label{dirac-pro}
\eeq
where
\beq
\mu(z)=\frac{s!!}{\Pf C}
\label{mu-def}
\eeq
and $s$ denotes de total number of second
class constraints. $\alpha,\beta=1,2...s$, $i,j,k=1,2,...n$ where $n$
is the dimension of the phase space and
\beq
\Pf C= \epsilon^{{\alpha_1}{\alpha_2}...{\alpha_{s-1}}{\alpha_{s}}}C_{\alpha_1
\alpha_2}... C_{\alpha_{s-1}\alpha_{s}},
\label{Pf}
\eeq
where
\beq
C_{\alpha_{i}\alpha_{j}}=\partial_l\chi_{\alpha_i}\sigma^{lk}\partial_k
\chi_{\alpha_j},
\eeq
is the Pfaffian associated to the matrix $C$ (\ref{C}). 

The central point here is that we can now give a constructive way to find the antisymmetric tensor $\eta^{iji_1...i_{s-1}i_s}$. 


That the structure (\ref{dirac-pro}) corresponds to  the  
Dirac bracket can be showed  by rewriting appropiately the original 
Dirac bracket structure defined in (\ref{dirac-sym}). 
As a first step we note that
\beq
{\sigma^*}^{ij}=-\frac1s
C^{\alpha\beta}\partial_k\chi_{\alpha}\partial_l\chi_{\beta}
\big(\sigma^{ij}\sigma^{kl}
+ s \sigma^{ik}\sigma^{lj}\big)
\eeq
by inserting a unity in the form
$C^{\alpha\gamma}C_{\gamma\beta}=\delta^\alpha_\beta$ in the first
term of (\ref{dirac-sym}).

Now, in general, the inverse of an antisymetric matrix $C$ can be
written in the form
\beq
C^{\alpha\beta}=-\frac{s}{\Pf
C}\epsilon^{\alpha\beta{\alpha_1}...{\alpha_{s-3}}{\alpha_{s-2}}}C_{\alpha_1
\alpha_2}... C_{\alpha_{s-3}\alpha_{s-2}}
\eeq
where $\Pf C$ is the Pfaffian, associated to the matrix $C$, defined in
(\ref{Pf}). 
With these notations the Dirac bracket can be rewritten in the form
\beq
{\sigma^*}^{ij}=\frac{1}{\Pf C}
\epsilon^{{\alpha_1}{\alpha_2}...{\alpha_{s}}}\partial_{i_1}\chi_{\alpha_1}
\partial_{i_2}\chi_{\alpha_2}....\partial_{i_{s-1}}\chi_{\alpha_{s-1}}
\partial_{i_s}\chi_{\alpha_s}
\omega^{iji_1i_2......i_{s-1}i_s}
\label{dirac-sigma}
\eeq
where
\beq
\omega^{iji_1i_2......i_{s-1}i_s}\equiv (\sigma^{ij}\sigma^{i_1i_2}+s
\sigma^{ii_i}\sigma^{i_2j})(\sigma^{i_3i_4}\sigma^{i_5i_6}....
\sigma^{i_{s-1}i_s})
\label{Sigma}
\eeq
In the expression (\ref{dirac-sigma}) 
the summation over the indices $\alpha_i$ can be
replaced by the sum of all the permutations over the indices of
(\ref{Sigma}), which comes from the Levi-Civita symbol. We can perform this sum in a constructive way. First in
the case $s=2$ we have
\beq
\label{s=2 definition}
\sum_P\omega^{iji_1i_2}=
2(\sigma^{ij}\sigma^{i_1i_2}-\sigma^{ii_1}\sigma^{ji_2}+
\sigma^{ii_2}\sigma^{ji_1})\equiv 2 \eta^{iji_1i_2}
\eeq 
Here $\eta^{iji_1i_2}$ is a definition of the tensor $\eta$ when $s=2$. In a recursive form, we can also construct the tensor $\eta$ in the case $s=4$ 
\bea
\nonumber
\sum_P\omega^{iji_1i_2i_3i_4}&=&4(\sigma^{ij}\eta^{i_1i_2i_3i_4}-
\sigma^{ii_1}\eta^{ji_2i_3i_4}+
\sigma^{ii_2}\eta^{ji_1i_3i_4}-
\sigma^{ii_3}\eta^{ji_1i_2i_4}\\
&+&\sigma^{ii_4}\eta^{ji_1i_2i_3})\equiv 8
\eta^{iji_1i_2i_3i_4}
\eea
with  the same definition for $\eta^{i_1i_2i_3i_4}$. In general
for arbitrary $s$ 
\bea
\nonumber
\sum_P\omega^{iji_1i_2...i_s}&=&s(\sigma^{ij}\eta^{i_1i_2..i_s}-
\sigma^{ii_1}\eta^{ji_2i_4...i_s}+
\sigma^{ii_2}\eta^{ji_1i_3...i_s}\\ \
&-&\sigma^{ii_3}\eta^{ji_1i_2i_4...i_s}+...
+\sigma^{ii_s}\eta^{ji_1i_2...i_{s-1}})\equiv s!!
\eta^{iji_1i_2...i_s}.
\label{epsi-def}
\eea
The Dirac structure has then the final form
\beq
{\sigma^*}^{ij}= \frac{s!!}{\Pf
C}\eta^{iji_1...i_{s-1}i_s}\partial_{i_1}\chi_{1}
\partial_{i_2}\chi_{2}...\partial_{i_{s-1}}\chi_{s-1}\partial_{i_s}\chi_{s}.
\label{Dirac-nuevo}
\eeq

For example when $n=4, s=2$, $\eta^{1324}=1$ and is $+1$ or $-1$
depending if the permutation of $1324$ is even or odd. As another
example when $n=6, s=2$
\beq
\eta^{1425}=1, \quad \eta^{1436}=1, \quad \eta^{2536}=1
\eeq
with its respective permutations. 

The DB in this new form has the following properties:
\begin{enumerate}

\item[(a)] Is analogue to the Hamiltonian construction outlined before for unconstrained dynamics
with $\mu(z)$ defined in (\ref{mu-def}). The second class
constraints are Casimir functions (\ref{prop1}) of the corresponding 
bracket but are
not integrals of motion of the corresponding dynamics.

\item[(b)] It fulfills the antisymmetry property (\ref{antisym}) by
construction and the Jacobi identity (\ref{jacobi}).

\item[(c)] Is invariant with respect to redefinitions of the constraint
surface: Under
the change of representation (\ref{changeM}) the structure
(\ref{Dirac-nuevo}) 
changes as
\beq
\bar{\sigma^*}^{ij}_p=\bar{\mu}(z)\eta^{iji_1...i_{s-1}i_s}
(M_1^{\alpha_1}....M_s^{\alpha_s})
(
\partial_{i_1}\chi_{\alpha_1}
\partial_{i_2}\chi_{\alpha_2}...\partial_{i_{s-1}}\chi_{\alpha_{s-1}}
\partial_{i_s}\chi_{\alpha_s}
)
\eeq
where $\bar\mu(z)=s!!/\Pf\bar C$ and $\bar C$ is the corresponding
transformed $C$ matrix.
This last expression can be written as
\beq
\bar{\sigma^*}^{ij}_p=\bar{\mu}(z)(\det M)\eta^{iji_1...i_{s-1}i_s}
(
\partial_{i_1}\chi_{1}
\partial_{i_2}\chi_{2}...\partial_{i_{s-1}}\chi_{s-1}\partial_{i_s}\chi_{s}
)
\eeq
and from (\ref{Dirac-nuevo}) we conclude that $\mu$ should transform as
\beq
\mu=\bar\mu \det M
\label{transmu}
\eeq
for $M$ an arbitrary matrix whose determinant is different from
zero. To show that indeed $\mu$ transform as (\ref{transmu}) we
observe 
that the  Pfaffian of the matrix $\bar C$
changes under the same transformation as
\beq
\Pf \bar C= (\det M) \Pf C.
\eeq
Therefore we conclude that the structure (\ref{Dirac-nuevo}) 
is invariant with respect to
redefinitions of the constraint surface.

\item[(d)] In a canonical
representation of the constraint surface 
defined by the canonical transformation
\beq
z^i\to (Q^\alpha, Q^r, P_\beta, P_r)
\eeq
where the constraint surface originally defined by $\chi_\alpha=0$ is now
represented in the new variables as $Q^\alpha=0, P_\beta=0$ where $
\{Q^\alpha,P_\beta\}=\delta^\alpha_\beta$, 
the function $\mu_{CR}(z)=1$. Indeed,
in these
variables  
the Dirac bracket takes the following canonical form
\beq
\{Q^{r}, P_s\}^*=\delta^r_s, 
\quad \{Q^\alpha, P_\beta\}^*=0.
\label{stand}
\eeq
By writing the structure (\ref{Dirac-nuevo}) in these new variables, we obtain
\beq
{\sigma^*}^{ij}_{CR}=\mu_{CR}\epsilon^{ij(1)(1+N)(2)(2+N)...(\alpha)(\alpha+N)}
\eeq
where we choose $Q^1=\chi_1, P_1=\chi_2...Q^\alpha=\chi_{s-1},
P_\alpha=\chi_s$ and here $\alpha,\beta=1,2,...s/2$.
As the Pfaffian in the canonical representation is $\Pf C_{CR}=s!!$ we
conclude that $\mu_{CR}=1$ as expected.
\end{enumerate}

\section{Dirac bracket as a Nambu structure}

In 1973 Nambu \cite{Nambu}  proposed a generalization of classical Hamiltonian
mechanics that can be applied to odd dimensional phase spaces. The
central idea is based on a generalization of the Poisson bracket
binary operation to a $n$-ary operation --the Nambu bracket--. The idea can be traced back to the work on n-ary Lie algebras of Filipov \cite{FP} where the fundamental identity that generalize the Jacobi identity was discovered.
Recent interest on this topic is due to \cite{Ho} where it is used in the context of M theory. Also the work \cite{Takhtajan} studied these generalizations of Poisson brackets with more than just two entries.
The consistence requirements on this generalization in an invariant
geometrical form similar to the one realized for the standard Hamiltonian 
mechanics was also developed in
\cite{Arnold}. These structures are  of relevance in Poisson-Lie group
structures \cite{CT}, generalization of Poisson structures \cite{AIPB},
integrable systems \cite{Hietarinta}, quantum
groups, quantization and deformation theory \cite{DFST}.  

As noticed in \cite{FP} and \cite{Takhtajan} this structure is more ``rigid'' than 
the Poisson
brackets in the sense that every Nambu structure can be recasted as a
Poisson bracket but not every Poisson bracket can be written as a Nambu
structure. In fact the generalization of the Jacobi identity for the
Poisson brackets --called in \cite{Takhtajan} 
the fundamental identity-- is much
more restrictive than the Jacobi identity. This is the reason why there
are relatively few known examples of these type of
structures (for some examples see \cite{Ch}, \cite{CT},
\cite{Hietarinta}). 
Nevertheless it is of interest to ask the question if the
Dirac bracket obtained in the previous section, that is a singular Poisson
structure, can be interpreted as a Nambu bracket. As we will see this question
can be solved in the affirmative for the case $n=s+2$ by constructing
the Dirac bracket as a subordinated structure of a Nambu bracket with $n$ 
entries where $s$ of them are fixed to be the second class constraints. In the general case, we can construct a Dirac-Nambu bracket which corresponds to a Nambu
bracket with $n$ entries fixing the same $s$ entries as in the previous case.
The resulting bracket is a Nambu bracket with $n-s$ entries as a 
subordinated structure. This result is in agreement with \cite{CZ} where it is proved in a different way for a specific version of a Nambu bracket defined using a determinant.

The Nambu structure is defined  by the following
properties \cite{Takhtajan}, \cite{FP} 

\begin{enumerate}
\item[(1)] Skew-symmetry
\beq
\{f_1,...f_n\}=(-1)^{e(\sigma)} \{f_{\sigma(1)}....f_{\sigma(n)}\}
\label{skewsym}
\eeq
where $\sigma$ is a permutation of $1...n$ and $e(\sigma)$ is its
parity.

\item[(2)] Leibniz rule

\beq
\{f_1f_2,f_3...f_{n+1}\}=f_1\{f_2,f_3...f_{n+1}\}+\{f_1,f_3...f_{n+1}\}f_2
\label{Leibniz}
\eeq

\item[(3)] Fundamental identity 
$$ \{\{f_1,...,f_n\},g_1,...,g_{n-1}\}$$ 
$$=  \{\{f_1,g_1,...,g_{n-1}\}, f_2,...,f_{n-1}\}+\{f_1 , \{f_2,g_1,...,g_{n-1}\},...,f_n\}  $$
\beq
+...+
\{f_1,...,f_{n-1},\{f_n,g_1,...,g_n\}\}
\label{FI}
\eeq
for any functions $f_1,...,f_n,g_1,...,g_{n-1}$ of the phase space
variables.
\end{enumerate}

Equations (\ref{skewsym}) and (\ref{Leibniz}) are the
standard 
skew-symmetry and
derivation properties found for the ordinary $(n=2)$ case Poisson
bracket, whereas (\ref{FI}) is a generalization of the Jacobi identity
that ensures the property that the Nambu bracket of two integrals of
motion is an integral of motion.

If we write the Nambu bracket in terms of the antisymetric tensor
$\eta$ \cite{Takhtajan}
\beq
\{f_1,...f_n\}=\eta^{1_1,...i_n}\partial_{i_1}f_1,...\partial_{i_n}f_n
\eeq
then the FI can be splited into two conditions \cite{Takhtajan},
\cite{Hietarinta}: the algebraic
condition 
\beq
{\cal N}^{i_1..i_nj_1...j_n}+{\cal N}^{j_1..i_ni_1,...j_n}=0
\label{alg}
\eeq
where
\bea
\nonumber
{\cal N}^{i_1..i_nj_1...j_n}&=&
\eta^{i_1i_2...i_n}\eta^{j_1j_2...j_n}+
\eta^{j_ni_1i_3...i_n}\eta^{j_1j_2...j_{n-1}i_2}+
\eta^{j_ni_2i_ii_4...i_n}\eta^{j_1j_2...j_{n-1}i_3}\\ 
&&+...+
\eta^{j_ni_2i_3...i_{n-1}i_n}\eta^{j_1j_2...j_{n-1}i_n}-
\eta^{j_ni_2i3...i_n}\eta^{j_1j_2...j_{n-1}i_1}
\eea
and the differential condition 
\bea
\nonumber
{\cal D}^{i_2...i_nj_1...j_n}&=&
\eta^{ki_2...i_n}\partial_k\eta^{j_1j_2...j_n}+
\eta^{j_nki_3...i_n}\partial_k\eta^{j_1j_2...j_{n-1}i_2}\\ \nonumber
&&+
\eta^{j_ni_2ki_4...i_n}\partial_k\eta^{j_1j_2...j_{n-1}i_3}
+...+
\eta^{j_ni_1i_3...i_{n-1}k}\partial_k\eta^{j_1j_2...j_{n-1}i_n}\\ 
&&-
\eta^{j_1j_2...j_{n-1}k}\partial_k\eta^{j_ni_2i_3...i_n}=0
\label{diff}
\eea
In general, it is very difficult to analyze the differential condition
for a given
Nambu tensor $\eta$ but we can take advantage of the analysis realized
in \cite{Hietarinta} that can be applied to any decomposable
antisymmetric tensor and reduces the differential condition to the
construction of some vector field and its commutation properties.

The Nambu bracket structure contains an infinite family of
``subordinated'' structures of lower degree that can be obtained from
the original structure by fixing some of its entries. We will show
that in some cases (depending on the dimensionality of the phase space
and the number of second class constraints) the Dirac bracket can be
retrived from a Nambu structure by fixing some of its entries to be
precisely the second class constraints.  

Let us start with the antisymmetric tensor with $s+2$ indices
\beq
\eta_*^{iji_1...i_s}= \mu(z) \eta^{iji_1...i_n}
\label{eta*}
\eeq
with $\eta$ and $\mu$ defined by (\ref{epsi-def}) and
(\ref{mu-def}) 
respectively. In the case $n=s+2$ the tensor $\eta$ coincides
with the antisymmetric Levi-Civita tensor with $s+2$ indices that run
over $1.....s+2$. The algebraic condition (\ref{alg}) is automatically
satisfied by ${\cal N}=0$. To prove that the differential condition
(\ref{diff} ) is also
satisfied we note that the tensor $\eta_*$ is decomposable which
means that can be written as a determinant (for details see
\cite{Hietarinta}). In that case ${\cal N}=0$
and the differential condition is scale invariant, i.e., for any
function $\rho(z)$ \cite{Hietarinta}
\beq
{\cal D}^{i_2...i_nj_1...j_n}(\rho\eta)=(\rho\partial_k\rho)
{\cal N}^{ki_2...i_nj_1...j_n}+\rho^2{\cal
D}^{i_2...i_nj_1...j_n}(\eta).
\eeq
In particular this means that if $\eta$ is a Nambu tensor $\rho\eta$
is also a Nambu tensor. Taking $\rho= \mu$ we then conclude that the differential
condition is also satisfied. This means that the tensor (\ref{eta*})
defines a Nambu bracket
\beq
\label{Nambu_definicion}
\{f_1,f_2,...,f_{s+2}\}=\eta_*^{iji_1....i_s}\partial_if_1\partial_jf_2...
\partial_{{i_s}}f_{s+2}.
\eeq
From this structure we can obtain the Dirac bracket
(\ref{Dirac-nuevo}) as a particular case by inserting
$f_3=\chi_1, f_4=\chi_2,...f_{s+2}=\chi_s$. Indeed
\beq\label{resultado}
\{f_1,f_2\}^*=\{f_1,f_2,\chi_1,...,\chi_s\}
\eeq
coincides with the Dirac bracket. It is interesting to note that the
Nambu structure is more general and ``proyects'' to the Dirac bracket
when some of its entries are fixed.

At first glance our result (\ref{resultado}) seems to conflict with the previous result of \cite{CZ} since it differs by a factor of $1/\{\chi_1,...,\chi_s\}$ (on the right hand side). However no contradiction arises, recall that our Nambu bracket definition is more general, specifically it differs from the usual one by a factor of $\mu (z)$ since we redefined it in (\ref{Dirac-nuevo}). But since $\mu (z)$ is the inverse of the Pfaffian of $C_{\alpha \beta}=\{\chi_\alpha,\chi_\beta \}$ (\ref{mu-def}), this terms accounts for the discrepancy using a relation between Nambu and Poisson brackets given in \cite{CZ}. Hence, both results agree. 

Generalizing our final result (\ref{resultado}), if the dimension of phase space is $n$ and $s$ is the number of
second class constraints we can define a generalized Dirac-Nambu bracket by
\beq \label{generalizacion}
\{f_1,f_2,...f_{n-s}\}^*=\{f_1,f_2,...,f_{n-s}, \chi_1,...,\chi_s\}
\eeq 
that is a Nambu structure with $n$ entries defined by
\beq \label{new_proposal}
\{f_1,f_2,...,f_{n-s}, \chi_1,...,\chi_s\}=\mu(z)\eta^{i_1...i_{n}}
\partial_{i_1}f_1\partial_{i_2}f_2....\partial_{i_{n}}\chi_{s}
\eeq
The generalization of the Nambu bracket to higher dimensions \cite{Chandre} where we have a bracket with more entries than just the Casimir functions plus 2 is also posible and can be used to propose this generalization of the Dirac-Nambu bracket for higher dimensions.

{\it Comment}: In the case  $n>s+2$,  the Dirac 
bracket defined in (\ref{Dirac-nuevo})
does not satisfy in general the conditions for a Nambu structure 
because the symplectic structure is not decomposable for $n>s+2$.
For example, in the case 
$n=6$, $s=2$ it is possible to show that the algebraic condition (\ref{alg}) is
not satisfied. Consider ${\cal N}_{12343566}$ which is equal to one but on
the other hand  ${\cal N}_{32341566}=0$.

\section{Examples}

\subsection{Two constraints}
As a particular example of the ideas exposed in this note take for
instance a phase
space defined by $z^i=(q_1,q_2,p_1,p_2)$ with the constraints
\beq
\chi_1=p_1+\frac12 B q_2 \quad \quad \chi_2=p_2-\frac12 B q_1
\eeq
So $n=4$ and $s=2$. The matrix $C^{\alpha\beta}$ has its Pffafian equal to $2B$. A Nambu bracket can be constructed as
\beq
\{F,G,\chi_1,\chi_2\}=\eta_{ijkl}^*\partial_iF\partial_jG\partial_k
\chi_1\partial_l\chi_2
\label{Namb}
\eeq
for $F(z),G(z)$ arbitrary functions of phase space variables,  
where the Nambu tensor $\eta_{ijkl}^*$ is
\beq
\eta_{ijkl}^*=\frac1B \eta_{ijkl}
\label{eta}=\frac1B \epsilon_{ijkl}
\label{eta}
\eeq
and $\epsilon$ is the Levi-Civita tensor in 4 dimensions. Since the dimesion of the phase space n is equal to the number of constraints s plus two, (\ref{Namb}) also defines a Dirac bracket using equation (\ref{Dirac-nuevo}), giving the result:
\beq \label{matriz ejemplo 1}
\sigma^*=\pmatrix{0&-\frac1B&\frac12&0\cr
                  \frac1B&0&0&\frac12\cr
                  -\frac12&0&0&-\frac{B}{4}\cr
                  0&-\frac12&\frac{B}{4}&0\cr}
\eeq

A direct application of the formula (\ref{Dirac-bra}) conduces to the same symplectic structure (\ref{matriz ejemplo 1}), thus showing the equivalence of both ways of calculating the Dirac bracket, as well as the equivalence (\ref{resultado}) to the Nambu bracket.

With the same constraints but in a higher dimensional phase space,
for example $z^i=(q_1,q_2,q_3,p_1,p_2,p_3)$ we can apply the same
formulas (\ref{Dirac-bra}) and (\ref{Dirac-nuevo}) to calculate the Dirac bracket and show that the results coincide:
\beq
\sigma^*=
\pmatrix{ 0 & -{1\over B} & 0 & {1\over 2} & 0 & 0 \cr 
{1\over B} & 0 & 0 & 0 & {1\over 2}
   & 0 \cr 
0 & 0 & 0 & 0 & 0 & 1 \cr 
-{1\over 2} & 0 & 0 & 0 & {{-B}\over 4} & 0 \cr 
0 &  -{1\over 2} & 0 & {B\over 4} & 0 & 0 \cr 
0 & 0 & -1 & 0 & 0 & 0 \cr  }.
\eeq
Nevertheless the corresponding proposal for the  Dirac bracket defined by (\ref{Namb}) and (\ref{eta}) fails to be a Nambu structure. Notice that in this case the tensor
$\eta_{ijkl}$ is not the Levi-Civita antisymmetric tensor but is
instead defined by (\ref{s=2 definition}) and (\ref{epsi-def}).

\subsection{Four constraints}
Now consider a phase space of $n=6$ dimensions  given by $z^i = (q_1, q_2, q_3, p_1, p_2, p_3)$, and s=4 constraints:
$$
\chi_1 = A_1 q_1 + B_2 p_2 \quad \quad \chi_2= B_1 p_1 - C_2 q_2 
$$
\begin{equation}
\chi_3= A_2 q_2 + C_3 p_3 \quad \quad \chi_4= B_3 p_2 + A_3 q_2
\end{equation}
where $A_1,A_2,A_3,B_1,B_2, B_3,C_2,C_3$ are constants.

A direct application of the standard Dirac bracket definition (\ref{Dirac-bra}) gives the following result:

\beq
\label{matriz_ej_2}
\sigma^*=
\pmatrix{ 
0 & 0 & \frac{A_3 B_2 C_3}{A_1 A_2 B_3} & 0 & 0 & 0 \cr 
0 & 0 & \frac{C_3}{A_2} & 0 & 0 & 0 \cr 
-\frac{A_3 B_2 C_3}{A_1 A_2 B_3} & -\frac{C_3}{A_2} & 0 & -\frac{C_2 C_3}{A_2 B_1} & \frac{A_3 C_3}{A_2 B_3} & 1 \cr 
0 & 0 & \frac{C_2 C_3}{A_2 B_1} & 0 & 0 & 0 \cr 
0 & 0 & -\frac{A_3 C_3}{A_2 B_3} & 0 & 0 & 0 \cr 
0 & 0 & -1 & 0 & 0 & 0 \cr  
}
\eeq

Now calculating the Dirac bracket with equation (\ref{Dirac-nuevo}), we get exactly the same result (\ref{matriz_ej_2}), just as in the first example. Moreover, since $n=s+2$ the structure is also equivalent to a Nambu bracket $\{F,G,\chi_1,\chi_2,\chi_3,\chi_4\}$ as in (\ref{Nambu_definicion}).

If we now use the same constraints but extend the phase space to $n=8$, $z^i=(q_1,q_2,q_3,q_4,p_1,p_2,p_3,p_4)$, in this case again both (\ref{Dirac-bra}) and (\ref{Dirac-nuevo}) give the same result:

\beq
\sigma^*=
\pmatrix{ 
0 & 0 & \frac{A_3 B_2 C_3}{A_1 A_2 B_3} & 0 & 0 & 0 & 0 & 0 \cr 
0 & 0 & \frac{C_3}{A_2} & 0 & 0 & 0 & 0 & 0\cr 
-\frac{A_3 B_2 C_3}{A_1 A_2 B_3} & -\frac{C_3}{A_2} & 0 & 0 & -\frac{C_2 C_3}{A_2 B_1} & \frac{A_3 C_3}{A_2 B_3} & 1  & 0 \cr 
0 & 0 & 0 & 0 & 0 & 0 & 0 & 1 \cr 
0 & 0 & \frac{C_2 C_3}{A_2 B_1} & 0 & 0 & 0 & 0 & 0 \cr 
0 & 0 & -\frac{A_3 C_3}{A_2 B_3} & 0 & 0 & 0 & 0 & 0 \cr 
0 & 0 & -1 & 0 & 0 & 0 & 0 & 0 \cr  
0 & 0 & 0 & -1 & 0 & 0 & 0 & 0 \cr  
}
\eeq

Which is almost the same matrix, only extended to the space $q_4,p_4$. However, now the Nambu bracket cannot be used to express this Dirac bracket. Instead, the only way to incorporate a Nambu bracket is with the proposed generalization (\ref{new_proposal}), using perhaps another Casimir function of the corresponding dynamics. 

\subsection{A particle on the surface $F(q^a)=0$}

A particle restricted to move on a surface defined by $F(q^a)=0$ with $a=1,...,d$ has
two constraints
\beq
\chi_1=F, \quad \chi_2=\sum_{m} z^{m+d}\partial_m F
\eeq
Notice that if $z=(q^a,p_a)$ and $p_a=\dot{q}^a$, it implies that $\chi_2=\mathbf{p}\cdot \nabla F=\frac{dF}{dt}$. A direct application of the formula (\ref{Dirac-nuevo}) conduces to
\beq
{\sigma^*}^{ij}=\frac{1}{|\nabla
F|^2}\eta^{ijkl}\partial_kF\partial_l(\sum_{m} z^{m+d}\partial_m F) 
\eeq
that coincides with the standard Dirac bracket calculated using equation (\ref{Dirac-bra}), which gives as a result:
\beq
\{q^i,q^j\}^*=0, \quad \{q^i,p_j\}^*=\delta^i_j-n_in_j, \quad
\{p_i,p_j\}^*= p_k(n_j\partial_kn_i-n_i\partial_kn_j)
\eeq
where $n_i=\partial_i F/|\nabla F|$. However, this only defines a Nambu bracket when d=2.

\subsection{Canonical representation}
As a final example consider a phase space of $n=6$ dimensions  given by $z^i = (q_1, q_2, q_3, p_1, p_2, p_3)$, and s=4 constraints:
\begin{equation}
    \chi_1=q_1 \quad \chi_2=p_1 \quad \chi_3=q_2 \quad \chi_4=p_2
\end{equation}

As in all previous examples, calculating the Dirac bracket using both (\ref{Dirac-bra}) and (\ref{Dirac-nuevo}) gives exactly the same result:
\begin{equation}
    \sigma^*=
\pmatrix{ 
0 & 0 & 0 & 0 & 0 & 0  \cr 
0 & 0 & 0 & 0 & 0 & 0  \cr 
0 & 0 & 0 & 0 & 0 & 1  \cr 
0 & 0 & 0 & 0 & 0 & 0  \cr 
0 & 0 & 0 & 0 & 0 & 0  \cr 
0 & 0 & -1 & 0 & 0 & 0 \cr  
}
\end{equation}
And again, since $n=s+2$ it can also be expressed using a Nambu bracket (\ref{Nambu_definicion}). But what is really interesting about this example is that the constraints are of the form studied in property (d) in section 4, that is, if  $\chi_1,...,\chi_s=q_1,p_1,...q_{s/2},p_{s/2}$, then $\Pf C=s!!$ and therefore $\mu=1$. In this example, using (\ref{Pf}) we calculate the Pfaffian, giving $\Pf C=8$. On the other hand, $s!!=4!!=(4)(2)=8$. Finally:
\begin{equation}
    \mu = \frac{s!!}{\Pf C}=\frac{8}{8}=1
\end{equation}
So this property in indeed verified.

\section{Conclusions}
In this work we gave an alternative formula for the Dirac bracket, and then a relation with the Nambu bracket. As we mentioned before, this is not entirely new as a formula was already derived in \cite{CZ} by a different approach, relying on a direct definition of the Nambu bracket using determinants. Here instead, we defined a Nambu bracket using only properties (\ref{skewsym})-(\ref{FI}) and proved the result relying only on its algebraic properties studied in \cite{Takhtajan} and \cite{Hietarinta}. We believe this is important as it allows more general definitions of a Nambu bracket to be related to a Dirac bracket, and therefore to the study of quantum systems. An interesting question for the future is the quantization of the dynamics defined by these more general Dirac-Nambu brackets (\ref{generalizacion}) as well as its potential applications to M-theory.

\section*{Acknowledgements}

Our work was partially supported by Mexico's National Council of Science and Technology (CONACyT) grant A1-S-22886 and DGAPA- UNAM grant IN116823. RCA was partially supported by a CONAHCyT (formerly CONACyT) Master's degree  scholarship 2023-000002-01NACF-08528 fellowship number 1272692.

\end{document}